\documentclass[aps,groupedaddress, superscriptaddress,showpacs,amsmath,amssymb,twocolumn,prl]{revtex4-1}
\usepackage{bm}
\usepackage[dvips]{graphicx}
\usepackage{wrapfig,subfigure}
\usepackage{amssymb}
\usepackage{color}
\usepackage[normalem]{ulem}
\usepackage{amsmath}
\usepackage{epstopdf}

\usepackage[pdfpagemode=UseNone,pdfstartview=FitH,colorlinks=true,linkcolor=blue,citecolor=blue,urlcolor=blue]{hyperref}
\usepackage[all]{hypcap}

\def\be{\begin{equation}}
\def\ee{\end{equation}}

\begin{document}
\title{Two-time correlators for propagating squeezed microwave in transients}
\author{Juan Atalaya}
\affiliation{Department of Electrical and Computer Engineering, University of California, Riverside, California 92521}
\author{Mostafa Khezri}
\affiliation{Department of Electrical and Computer Engineering, University of California, Riverside, California 92521}
\affiliation{Department of Physics, University of California, Riverside, California 92521}
\author{Alexander N.\ Korotkov}
\affiliation{Department of Electrical and Computer Engineering, University of California, Riverside, California 92521}
\date{\today}

\begin{abstract}
We analyze two-time correlators as the most natural characteristic of a propagating quadrature-squeezed field in the transient regime. The considered system is a parametrically driven resonator with a time-dependent drive. Using a semiclassical approach derived from the input-output theory, we develop a technique for calculation of the two-time correlators, which are directly related to fluctuations of the measured integrated signal. While in the steady state the correlators are determined by three parameters (as for the phase-space ellipse describing a squeezed state), four parameters are necessary in the transient regime. The formalism can be generalized to weakly nonlinear resonators with additional coherent drive. We focus on squeezed microwave fields relevant to the measurement of superconducting qubits; however, our formalism is also applicable to optical systems. The results can be readily verified experimentally.
\end{abstract}
\maketitle

Squeezed microwave fields (SMFs) \cite{Movshovich1990} have recently become the focus of extensive research efforts, related to superconducting quantum computing. This was enabled by a rapid progress in the development of practical superconducting parametric amplifiers \cite{Castellanos2008, Bergeal2010, Vijay2011, Tsai2008, Mutus2013}, which have become versatile sources as well as detectors of SMFs. Applications of intracavity and propagating (itinerant) SMFs include qubit readout \cite{Didier2015, Eddins2018}, metrology \cite{Caves1981, Lehnert2016, Berlet2017}, continuous-variable entanglement \cite{Eichler2011, Flurin2012}, control of artificial-atom fluorescence \cite{Toyli2016}, etc. Among other experimental achievements are demonstrations of the dynamic Casimir effect \cite{Wilson2011, Hakonen2013},  tomography of an itinerant SMF \cite{Mallet2011}, and detection of SMF radiation pressure \cite{Clark2016}.

Besides generation in phase-sensitive parametric amplification, SMFs are also self-generated in the process of circuit QED measurement of superconducting qubits \cite{Sete2013, Khezri2016} due to effective nonlinearity of the resonator induced by coupling with the qubit. Since squeezing affects the qubit measurement error, and for fast readout
the steady-state regime is not reached, analysis of squeezing in transients is very important. The corresponding dynamics of the intracavity squeezing has been recently analyzed \cite{Khezri2017}; however, there is still no theory for transient squeezing of the propagating SMF, which determines the qubit measurement accuracy. Moreover, our extensive search for any papers discussing transient evolution for a resonator-produced propagating squeezed field resulted in only a few remotely related references \cite{Ekert1989, Yurke1987, Ficek1984, Stassi2016}, which cannot serve as a starting point in developing a theory to answer this physically interesting and practically important question.

In this work, we analyze the {\it transient regime} of the propagating SMF, generated by a parametrically-driven linear resonator \cite{Dykman2012}, as shown in Fig.\ 1(a). The case of a weakly nonlinear resonator with a coherent drive (as in the qubit measurement) is slightly more complicated but equivalent, as discussed in the Supplemental Material \cite{SM}. As needed for practical applications, we focus on two-time correlators \cite{Dykman1975} for the quadrature (homodyne) measurement \cite{Milburn-Walls, Bozyigit2011}, with quadrature angle $\varphi$ changing in time. In particular, we find that in transients the dependence of the correlator on two angles $\varphi_1$ and $\varphi_2$ is characterized by four parameters, in contrast to only three parameters needed in a steady state, as for the  ellipse in phase space, which is traditionally used to describe squeezing. Our results can be readily checked experimentally.

\begin{figure}[tb!]
\centering
\includegraphics[width=0.9\linewidth, trim = 7.0cm 3cm 5.7cm 2.5cm,clip=true]{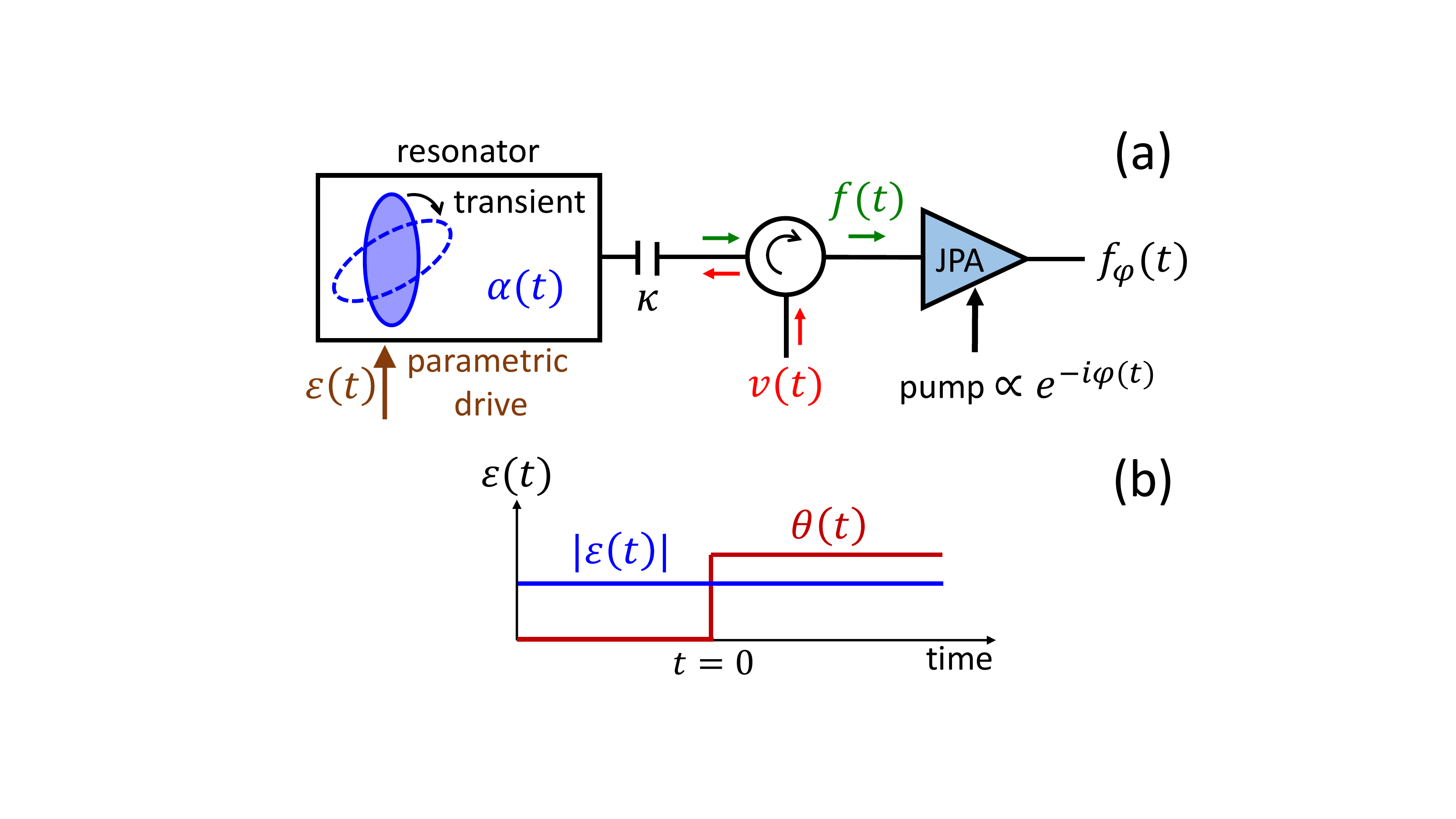}
\caption{(a) Analyzed system. Propagating microwave field [described by operator $F(t)$ or complex stochastic variable $f(t)$] is squeezed due to parametric drive of the resonator with changing in time amplitude $\varepsilon(t) = |\varepsilon(t)| \, e^{i\theta(t)}$. The amplified quadrature phase $\varphi (t)$ also changes in time, producing the noisy output signal $f_\varphi (t)$. The resonator damping rate is $\kappa$, and the incoming vacuum noise is described by $v (t)$. (b) An example of the parametric drive change, producing transient evolution of the resonator field [depicted in panel (a)] and of the propagating field.
}
\label{fig:setup}
\end{figure}

{\it System and Hamiltonian.}  Let us consider a parametrically modulated resonator [Fig.\ 1(a)] described in the rotating-wave approximation by the Hamiltonian ($\hbar=1$)
    \begin{align}
H = \Omega (t) \, a^\dagger a + \frac{i}{4} \left[ {\varepsilon^*(t) \, {a}^2 - \varepsilon (t)\, a^\dagger}^2 \right],
    \label{eq:Hamiltonian}\end{align}
where the resonator detuning  $\Omega (t)=\omega_{\rm r}(t)-\omega_{\rm d}$  and the parametric drive amplitude $\varepsilon (t) = |\varepsilon (t)| \, e^{i\theta (t)}$ can depend on time (slowly in comparison with the rotating frame frequency $\omega_{\rm d}$). In the laboratory frame, this Hamiltonian corresponds to the resonator frequency modulation at the double-frequency, $\omega_{\rm r}- |\varepsilon| \sin (2\omega_{\rm d}t-\theta)$. The more general case of a nonlinear resonator and added  coherent drive is discussed in the Supplemental Material \cite{SM}.

The propagating microwave field leaking from the resonator, described by operator $F(t)$, is amplified by a {\it phase-sensitive} amplifier, which amplifies the quadrature phase $\varphi$, so that the measured operator  is $F_\varphi(t)  = [F(t)\, e^{-i\varphi}+ F^\dagger (t)\, e^{i\varphi}]/2$. In contrast to most previous works, we assume a time-dependent phase $\varphi(t)$. After the mixer [not shown in Fig.\ 1(a)], the $\varphi$-quadrature measurement produces a {\it classical} (normalized) fluctuating output signal $f_\varphi (t)$, which in a typical experiment is integrated with a weight function $w(t)$ to produce the measurement result $R=\int w(t) f_\varphi (t)\, dt$. To analyze fluctuations of $R$, we need $\langle R^2\rangle=\iint w(t_1) w(t_2)\langle f_{\varphi_1} (t_1 ) f_{\varphi_2} (t_2)\rangle \, dt_1 dt_2$, where $\varphi_k\equiv \varphi(t_k)$. Therefore, in experiments it is important to know the correlator
    \be
    K_{\varphi_1\varphi_2}(t_1, t_2)\equiv \langle f_{\varphi_1}(t_1) \, f_{\varphi_2}(t_2)\rangle ,
    \label{eq:correlator-def}\ee
which will be the main object analyzed in this paper. Note that in our model, $f_\varphi (t)$ is only noise (amplified and measured propagating squeezed vacuum), i.e.\ $\langle f_\varphi (t)\rangle=0$; it is simple to add a non-zero signal by adding a coherent drive \cite{SM} into Eq.\ (\ref{eq:Hamiltonian}), but this does not affect fluctuations because of linearity.
For simplicity, we assume that the resonator energy decay rate $\kappa$ is only due to coupling $\kappa_{\rm out}$ with the transmission line, $\kappa =\kappa_{\rm out}$ (generalization to the case $\kappa >\kappa_{\rm out}$ is trivial in the same-temperature case, see below).

In the simplest case of zero detuning ($\Omega=0$), zero temperature,  and time-independent $\varphi$ and $\varepsilon$, the propagating squeezed vacuum produces the steady-state correlator
    \begin{align}
& K_{\varphi\varphi}(0, \tau) = \frac{\delta (\tau)}{4} -\frac{\kappa |\varepsilon|}{4\kappa_+} \, e^{-\kappa_+ |\tau|/2} \cos^2(\varphi-\theta /2)
    \nonumber \\
& \hspace{0.7cm}+  \frac{\kappa |\varepsilon|}{4 \kappa_-} \, e^{-\kappa_-|\tau|/2} \sin^2(\varphi-\theta /2), \,\,\, \kappa_\pm =\kappa \pm |\varepsilon|,
    \label{corr-simple}\end{align}
as can be obtained via the conventional input-output formalism \cite{GardinerBook, Clerk2010}, assuming $|\varepsilon|<\kappa$. Correspondingly, the integrated correlator for $\varphi=\theta/2$ is $\int_{-\infty}^\infty K_{\varphi\varphi}(0, \tau)\, d\tau =(1/4)(\kappa_-/\kappa_+)^2$, so it is squeezed compared with the  vacuum value of $1/4$, while for $\varphi=(\theta+\pi)/2$ it is unsqueezed: $\int_{-\infty}^\infty K_{\varphi\varphi}(0, \tau)\, d\tau =(1/4)(\kappa_+/\kappa_-)^2$.

Note that dependence of the correlator $K_{\varphi\varphi}(0,\tau)$ on $\varphi$ is described by three real parameters. Also note that since in the steady state $K_{\varphi\varphi}(0,\tau)$ depends only on the time difference $\tau\equiv t_2-t_1$, it is natural to use the Fourier transform, so the squeezing is usually analyzed in terms of the squeezing spectrum \cite{Milburn-Walls, Drummond-book}  $S_\varphi(\omega)\equiv 4 \int_{-\infty}^\infty e^{-i\omega \tau} K_{\varphi\varphi}(0,\tau)\, d\tau$. However, during transients such a Fourier transform is not natural, so we will focus on the two-time correlator $K_{\varphi_1\varphi_2}(t_1, t_2)$.

{\it Semiclassical model for measured fluctuations.} Instead of using the conventional input-output formalism~\cite{GardinerBook}, we will use a  simpler semiclassical stochastic model \cite{CarmichaelBook} to analyze the temporal correlations of the output signal $f_\varphi(t)$. As shown in \cite{SM}, the correlators obtained using this model are {\it exact} for our linear system (\ref{eq:Hamiltonian}); the model is still a good approximation for a weakly nonlinear resonator.

In this semiclassical model, the fluctuation of the (quantum) propagating output field $F(t)$ is treated as a complex-valued stochastic variable,
\begin{align}
\label{eq:f-def}
f(t) = -v  (t) + \sqrt{\kappa}\, \alpha(t),
\end{align}
where the complex-valued stochastic variable $\alpha(t)$ describes fluctuations of the intracavity field, while the incoming vacuum noise [Fig.\ 1(a)] is described by  a complex-valued Gaussian noise $v  (t)$ with two-time correlators
\begin{align}
\label{eq:noise-corr}
\hspace{-0.1cm} \langle v (t)\, v ^*(t')\rangle = (\bar{n}_{\rm b}+1/2)\,\delta (t-t'), \,\,\, \langle v (t)\, v (t')\rangle =0,
\end{align}
where $\langle ... \rangle$ denotes ensemble average and $\bar{n}_{\rm b}=[\exp(\omega_{\rm r}/T)-1]^{-1}$ is the average number of bath thermal photons. For brevity of formulas, we will assume the temperature $T$ to be zero (so $\bar{n}_{\rm b}=0$); however, for $T\neq 0$ all correlators in this paper can be simply multiplied by the factor $1+2\bar{n}_{\rm b}$.

The intracavity field fluctuation $\alpha(t)$ for a parametrically modulated resonator (\ref{eq:Hamiltonian}) evolves as
\begin{align}
\label{eq:eom-alpha}
\dot\alpha(t) =&\, -\left(\frac{\kappa}{2} + i\Omega \right)\alpha(t) - \frac{\varepsilon(t)}{2}\, \alpha^*(t) + \sqrt{\kappa}\, v (t) .
\end{align}
Note that in our normalization, $|\alpha|^2$ corresponds to the number of photons in the resonator, while $|f|^2$ corresponds to the propagating number of photons per second. The decay rate $\kappa$ is frequency-independent, i.e., we use the Markovian approximation \cite{Dykman1975}. The term $-\varepsilon\alpha^*/2$ describes effective increase of $\kappa$ by $|\varepsilon|$ for the quadrature phase $\varphi=\theta /2$ and its decrease by $|\varepsilon|$ for $\varphi=(\theta +\pi)/2$.

The output signal $f_\varphi(t)$ from the quadrature measurement is given by the real-valued stochastic variable
    \be
    f_\varphi (t) = {\rm Re}[e^{-i\varphi(t)} f(t)],
    \ee
so the correlator of interest (\ref{eq:correlator-def}) can be calculated as
\begin{align}
\label{eq:K-varphi}
&K_{\varphi_1\varphi_2}(t_1,t_2) = \frac{1}{2} \, {\rm Re}\left[K_{ff}(t_1,t_2) \, e^{-i(\varphi_1 + \varphi_2)}\right]
    \nonumber \\
& \hspace{2.3cm} + \frac{1}{2} \, {\rm Re}\left[K_{f f^{^*}}(t_1,t_2) \, e^{-i(\varphi_1 - \varphi_2)} \right] ,
    \\
& K_{ff}(t_1,t_2) \equiv \langle f(t_1) \, f(t_2)\rangle,
    \label{eq:Kff}\\
& K_{f f^{^*}}(t_1,t_2) \equiv \langle f(t_1) \, f^*(t_2)\rangle.
    \label{eq:Kff*}\end{align}

We see that for given $t_1$ and $t_2$, the dependence of $K_{\varphi_1\varphi_2}(t_1,t_2)$ on $\varphi_1$ and $\varphi_2$ is described by {\it four real parameters} [e.g., ${\rm Re}(K_{ff})$, ${\rm Im}(K_{ff})$, ${\rm Re}(K_{ff^*})$, and ${\rm Im}(K_{ff^*})$]. As will be discussed later, in the steady state there are only three independent real parameters because $K_{ff^*}$ in this case is real. Note that $K_{ff}$ and $K_{ff^*}$ obviously satisfy the symmetry relations \cite{Mandel1987},
\begin{align}
\label{eq:Kff-Kff*-prop}
\hspace{-0.1cm}K_{ff}(t,t')=K_{ff}(t',t), \,\,\,  K_{ff^{^*}}(t,t')=[K_{ff^{^*}}(t',t)]^*.
\end{align}

Now let us calculate the correlators $K_{ff}(t_1, t_2)$ and $K_{ff^*}(t_1, t_2)$ using the semiclassical model (\ref{eq:f-def})--(\ref{eq:eom-alpha}). Because of the symmetry, it is sufficient to assume $t_2>t_1$ (the $\delta$-function contribution to $K_{ff^*}$ at $t_1=t_2$ is discussed below). Let us introduce the column vector containing both correlators, $\mathbf{K}(t_1,t_2)=\big( K_{ff}(t_1,t_2),\, K_{ff^{^*}}(t_1,t_2)\big)^{\rm T}$.
From Eq.~\eqref{eq:f-def} we obtain
\begin{equation}
\label{eq:deriv-1}
\mathbf{K}(t_1,t_2)=\kappa  \left[ {\begin{array}{c} \langle\alpha(t_2)\, \alpha(t_1)\rangle  \\ \langle\alpha^*(t_2) \, \alpha(t_1)\rangle \end{array}}\right] -\sqrt{\kappa}\left[ {\begin{array}{c} \langle\alpha(t_2)\, v (t_1)\rangle  \\ \langle \alpha^*(t_2)\, v (t_1)\rangle \end{array}}\right],
\end{equation}
since $\langle  v (t_2) \, \alpha(t_1) \rangle=\langle v ^*(t_2) \, \alpha(t_1) \rangle=0$ because of causality.
Now using Eq.~\eqref{eq:eom-alpha}, we find the evolution of $\mathbf{K}(t_1,t_2)$ as a function of $t_2$,
\begin{align}
\label{eq:K-vector}
\partial {\mathbf{K}}(t_1,t_2)/\partial t_2 = M(t_2) \, {\mathbf{K}}(t_1,t_2),
\end{align}
where the matrix $M(t)$ describes the ensemble-averaged evolution of the vector $(\alpha,\alpha^*)^{\rm T}$ following from Eq.\  \eqref{eq:eom-alpha} without the noise term (contribution from the noise $v $ averages to zero because of linearity),
\begin{align}
\label{eq:M-ens}
M(t) = \left[ {\begin{array}{cc} -\kappa/2-i\Omega & -\varepsilon(t)/2
\\
-\varepsilon^*(t)/2 &  -\kappa/2+i\Omega \end{array}}\right].
\end{align}
Note that $M(t)$ is Hermitian only if $\Omega=0$.

To find the initial condition for Eq.~\eqref{eq:K-vector} at $t_2=t_1+0$, we use Eq.\ (\ref{eq:deriv-1}) with $\langle\alpha (t_1+0)\,v (t_1)\rangle =0$ and $\langle\alpha^* (t_1+0)\,v (t_1)\rangle =\sqrt{\kappa}/2$, where the last equation follows from Eq.\ \eqref{eq:eom-alpha}: $\alpha^*(t_1+dt) \approx \alpha^*(t_1) + \sqrt{\kappa} \, v ^*(t_1)\,dt$, while $\langle |v  (t_1)|^2\rangle =1/(2\,dt)$ from Eq.~\eqref{eq:noise-corr}. Therefore,
\begin{align}
\label{eq:K-vector-initial}
\mathbf{K}(t_1,t_1+0) = \kappa\left[ {\begin{array}{c} \langle\alpha^2(t_1)\rangle  \\ \langle|\alpha^2(t_1)|\rangle - 1/2\end{array} }\right].
\end{align}

The solution of Eq.\ (\ref{eq:K-vector}) with the initial condition (\ref{eq:K-vector-initial}) can be expressed via the Green's function $2\times2$ matrix $G(t |t_{\rm in})$, defined as
    \be
\partial G(t|t_{\rm in})/\partial t = M(t)\, G(t|t_{\rm in}), \,\,\, G(t_{\rm in}|t_{\rm in})=\openone.
    \label{eq:Green}\ee
Thus, for $\mathbf{K}$ (now expressed via $K_{ff}$ and $K_{ff^*}$) we obtain
\begin{equation}
\label{eq:deriv-2}
\left[ {\begin{array}{c} K_{ff}(t_1,t_2) \\ K_{f f^{^*}}(t_1,t_2)\end{array}}\right] = \kappa \, G(t_2|t_1) \left[ {\begin{array}{c} \langle\alpha^2(t_1)\rangle  \\ \langle|\alpha^2(t_1)|\rangle - 1/2 \end{array} }\right].
\end{equation}

To complete the calculation of $K_{ff}$ and $K_{ff^*}$, we need the second moments of the intracavity field fluctuations, $\langle \alpha^2(t_1)\rangle$ and $\langle |\alpha^2(t_1)|\rangle$. Following the result of Ref.\ \cite{Khezri2017}, they can be obtained as a solution of a system of four first-order differential equations. Alternatively, they can be obtained from Eq.~\eqref{eq:eom-alpha} as (see \cite{Ludwig1975})
\begin{align}
\label{eq:moments-t}
&\hspace{-0.cm}\left[ \begin{array}{cc}
 \langle|\alpha^2(t_1)|\rangle &  \langle\alpha^2(t_1)\rangle  \\
\langle {{\alpha}^*}^2(t_1)\rangle&  \langle|\alpha^2 (t_1)|\rangle
\end{array}
\right] = \frac{\kappa}{2} \int_{t_0}^{t_1} G(t_1 |t')\, G^\dagger(t_1 | t')\, dt'
    \nonumber \\
&\hspace{0.8cm} + G(t_1 | t_0)\left[ \begin{array}{cc}
 \langle|\alpha^2(t_0)|\rangle &  \langle\alpha^2(t_0)\rangle  \\
\langle {{\alpha}^*}^2(t_0)\rangle&  \langle|\alpha^2 (t_0)|\rangle
\end{array}
\right] G^\dagger(t_1 | t_0),
\end{align}
where $\langle \alpha^2(t_0)\rangle = {\rm Tr}[a^2\rho(t_0)]$, $\langle |\alpha^2(t_0)|\rangle = {\rm Tr}[a^\dagger a\, \rho(t_0)] +1/2$, and  $\rho (t_0)$ is a given intracavity state at an initial time $t_0$  (for $t_0\to -\infty$, the initial state is irrelevant).

Equations (\ref{eq:Green})--(\ref{eq:moments-t}) are the main result of this paper. Using these equations with $M(t)$ defined in Eq.\ (\ref{eq:M-ens}), we can find the correlators $K_{ff}$ and $K_{ff^*}$, which can then be used to obtain the main correlator of interest $K_{\varphi_1\varphi_2}(t_1,t_2)$ via Eq.\ (\ref{eq:K-varphi}). As mentioned above, in the case of a non-zero bath temperature, the correlators should be multiplied by $1+2\bar{n}_{\rm b}$.

At $t_2=t_1$, the correlator $K_{ff^*}$ contains the singular contribution $(\bar{n}_{\rm b}+1/2)\,\delta(t_2-t_1)$, as follows from Eqs.\ (\ref{eq:f-def}) and (\ref{eq:noise-corr}), while $K_{ff}$ does not have a singularity. Since in this case $\varphi_1=\varphi_2$, the correlator $K_{\varphi_1\varphi_2}(t_1,t_2)$ has the singular contribution $(1/4)(1+2 \bar{n}_{\rm b})\,\delta(t_2-t_1)$. In a real experiment, at $t_2\approx t_1$ there is also a contribution from the additional noise of a not-quantum-limited amplifier.

In the derivation we assumed that energy decay in the resonator is only due to coupling with the outgoing transmission line, i.e.\ $\kappa=\kappa_{\rm out}$. If this is not the case, the correlators $K_{\varphi_1\varphi_2}$, $K_{ff}$, and $K_{ff^*}$ for $t_1\neq t_2$  should be simply multiplied by the factor $\kappa_{\rm out}/\kappa$. This can be shown by
repeating the derivation with Eq.\ (\ref{eq:f-def}) replaced by  $f=-v+\sqrt{\kappa_{\rm out}}\, \alpha$ and Eq.\ (\ref{eq:eom-alpha}) replaced by $\dot{\alpha}=-(\kappa/2+i\Omega )\alpha -(\varepsilon/2)\alpha^* +\sqrt{\kappa_{\rm out}}\, v +\sqrt{\kappa-\kappa_{\rm out}}\, v_{\rm add}$, where the additional uncorrelated noise $v_{\rm add}(t)$ satisfies Eq.\ (\ref{eq:noise-corr}) with the same temperature. Alternatively, the multiplication of the correlators by $\kappa_{\rm out}/\kappa$ is rather obvious because the system is then equivalent to adding a beamsplitter with transmission amplitude $\sqrt{\kappa_{\rm out}/\kappa}$ to the outgoing transmission line (after the circulator) in Fig.\ 1(a). Note that the singularity of correlators at $t_2=t_1$ does not change when $\kappa_{\rm out}\neq \kappa$, because of the additional noise.

Even though our results have been derived for the case of a linear parametrically-driven resonator (\ref{eq:Hamiltonian}), we emphasize that they remain practically the same if a weak nonlinearity is added to the  resonator, as well as a coherent drive (see \cite{SM}). In this case the evolution of fluctuations should be linearized in the vicinity of the classical evolution (this modifies the matrix $M$) and we need to use the Gaussian approximation.

{\it Steady-state regime.} In the steady state we can  assume that the parametric drive amplitude $\varepsilon$ does not depend on time (as well as parameters $\Omega$ and $\kappa$). This is the case considered in the literature (e.g., \cite{GardinerBook,CarmichaelBook,Milburn-Walls}). Using our formalism (with $\bar{n}_{\rm b}=0$), we can easily find the Green's function $G(t|t_{\rm in})$ by finding eigenvalues and eigenvectors of the matrix $M$. Then from Eqs.\ (\ref{eq:deriv-2}) and (\ref{eq:moments-t}) we obtain
    \begin{align}
& K_{ff}(0,\tau) = - \frac{\kappa\varepsilon}{4} \left[ \left( 1-\frac{2i\Omega}{\epsilon}\right) \frac{e^{-\kappa_- |\tau|/2}}{\kappa_-} \right.
    \nonumber\\
& \hspace{2.3cm}  \left.  +
\left( 1+\frac{2i\Omega}{\epsilon}\right) \frac{e^{-\kappa_+ |\tau|/2}}{\kappa_+} \right] ,
    \label{eq:steady-ff}\\
& K_{ff^*}(0,\tau)= \frac{\delta (\tau)}{2}+ \frac{\kappa |\varepsilon|^2}{4\epsilon}
 \left( \frac{e^{-\kappa_- |\tau|/2}}{\kappa_-} - \frac{e^{-\kappa_+ |\tau|/2}}{\kappa_+} \right) ,
    \label{eq:steady-ff*}\end{align}
where $\kappa_\pm=\kappa\pm \epsilon$ and $\epsilon =\sqrt{|\varepsilon|^2-4\Omega^2}$ if $|\Omega|<|\varepsilon|/2$ (overdamped case) or $\epsilon =i\sqrt{4\Omega^2-|\varepsilon|^2}$ if $|\Omega|>|\varepsilon|/2$ (underdamped case). The condition of  stability is obviously $|\varepsilon|^2<\kappa^2+4\Omega^2$. The singular contribution $\delta (\tau)/2$ added into Eq.\ (\ref{eq:steady-ff*}) follows from Eqs.\ (\ref{eq:f-def}) and (\ref{eq:noise-corr}).

We see that in the steady state, $K_{ff^*}(0,\tau)$ is always real. Therefore, the squeezing is determined by three real parameters (which depend on $\tau$), in contrast to four parameters in the general (transient) case.

A convenient way of introducing the four  real parameters ($A$, $B$, $\phi$ and $\psi$) is by rewriting Eq.\ (\ref{eq:K-varphi}) as
    \begin{align}
& K_{\varphi_1\varphi_2}(t_1,t_2) = A\cos(\varphi_1-\phi)\,\cos(\varphi_2-\psi)
\nonumber \\
&\hspace{0.8cm} + B\sin(\varphi_1-\phi)\,\sin(\varphi_2-\psi) + \delta(t_2-t_1)/4, \label{eq:corr-param}
    \end{align}
where we explicitly added the singular term (note that $\varphi_1=\varphi_2$ when $t_1=t_2$) and the parameters $A$, $B$, $\phi$ and $\psi$ (all depending on $t_1$ and $t_2$) can be obtained from equations $(A-B)\,e^{i(\phi+\psi)}= K_{ff}(t_1,t_2)$ and  $(A+B)\,e^{i(\phi-\psi)}= K_{ff^*}(t_1,t_2)-\delta(t_2-t_1)/2$. As discussed above, in the steady state $K_{ff^*}$ is real, and therefore $\phi=\psi$, thus again leaving only three independent real parameters.

Note that in the case when $\varphi_1=\varphi_2$, the correlator $K_{\varphi\varphi} (t_1, t_2)$ drawn in the phase space as a function of the polar angle $\varphi$ is always an ellipse (even in the transient regime), as follows from Eq.\ (\ref{eq:corr-param}). In the steady state, from the measured three parameters of this ellipse it is possible to find all parameters in Eq.\ (\ref{eq:corr-param}) ($A$, $B$, and $\phi=\psi$), thus predicting the correlator for $\varphi_1\neq \varphi_2$ as well. However, in the general (transient) case this is impossible because of one extra parameter.

{\it Example of transient evolution.} To observe experimentally the discussed features of the squeezing in transients, the simplest case is to use no detuning ($\Omega =0$) and to change abruptly the parametric drive amplitude $|\varepsilon(t)|\, e^{i\theta (t)}$ (with a reasonably long cycle to accumulate ensemble statistics). If only $|\varepsilon (t)|$ is changing \cite{Ekert1989}, then the dynamics is still not very interesting (squeezing is still characterized by only three parameters). Therefore, the natural choice is to keep $|\varepsilon|$ constant, but to change abruptly the phase $\theta (t)$, as shown in Fig.\ 1(b). Let us assume that $\theta(t)=0$ for $t<0$ and $\theta (t)=\tilde\theta$ for $t>0$. Then solving Eqs.\  (\ref{eq:Green})--(\ref{eq:moments-t}) we obtain

\begin{align}
&\hspace{1.5cm}  K_{ff}(t_1,t_1+\tau) = \kappa [P_-+P_+]\, e^{i\tilde\theta},
    \\
&\hspace{1.5cm}  K_{ff^{^*}}(t_1,t_1+\tau) = \kappa[P_+ -P_-],
    \\
& P_\pm =\Big\{ \frac{\kappa |\varepsilon|} {4(\kappa^2 -|\varepsilon|^2)}\big[(1-\cos\tilde\theta) \, e^{-\kappa_\pm t_1} + i\sin\tilde\theta \, e^{-\kappa t_1}\big]
\nonumber \\
& \hspace{1.2cm}
- \frac{|\varepsilon|}{4\kappa_\pm} \Big\}\, e^{-\kappa_\pm \tau/2}, \,\,\,
\end{align}
where $\kappa_\pm =\kappa \pm |\varepsilon|$ and $\tau>0$.
Figure 2 shows the corresponding parameters $A$, $B$, $\phi$ and $\psi$ in Eq.\ (\ref{eq:corr-param}) as functions of $\tau$ for several values of $t_1$. As expected, we see that $\phi \neq \psi$, except in the steady state ($t_1\to \infty$).

\begin{figure}[t!]
\centering
\includegraphics[width=0.87\linewidth, trim = 1.5cm 1cm 1cm 1cm,clip=true]{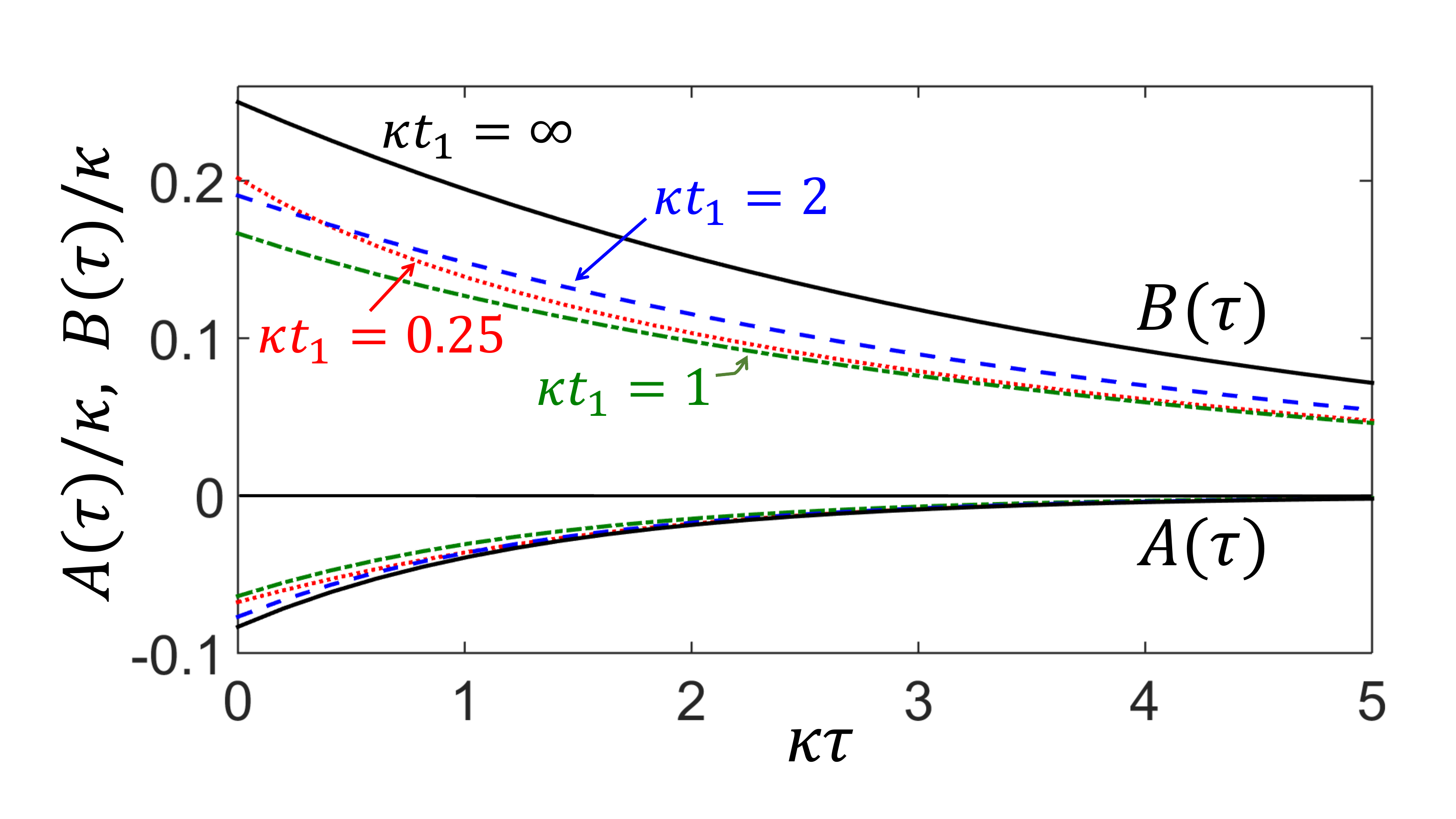}
\includegraphics[width=0.87\linewidth, trim = 1.5cm 1cm 1cm 1cm,clip=true]{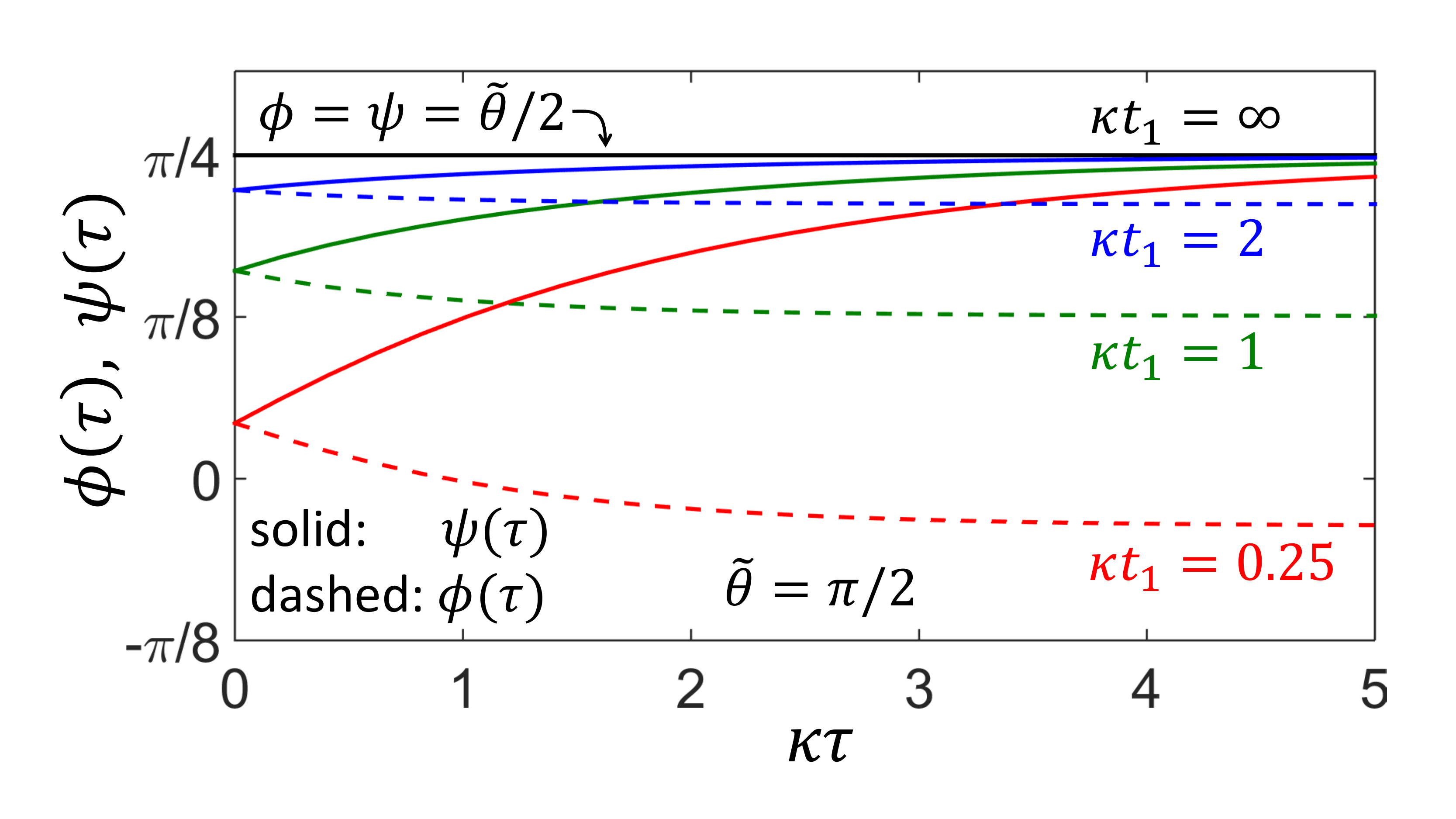}
\caption{Parameters $A$ and $B$ (top panel) and $\phi$ and $\psi$ (bottom panel) as functions of $\tau=t_2-t_1$ for several values of time $t_1$ passed after the abrupt change of the parametric drive shown in Fig. 1(b), $\kappa t_1=0.25,\, 1,\, 2, \infty$. In the steady state, $\phi=\psi$. We use $\tilde\theta=\pi/2$,  $|\varepsilon|/\kappa=0.5$, and $\Omega=0$. }
\label{fig:fig-transient}
\end{figure}

Thus, in this example the steady-state squeezing is described by three parameters; $A(\tau)$, $B(\tau)$, and $\phi$ (not depending on $\tau$), while the transient squeezing is described by four parameters: $A$, $B$, $\phi$, and $\psi$, which all depend on both $\tau$ and $t_1$. The same conclusion of three versus four parameters remains true if the correlator $K_{\varphi_1\varphi_2}(t_1, t_1+\tau)$ is integrated over $\tau$ or if we apply the Fourier transform over $\tau$ (as in the squeezing spectrum).

Note that to check our results experimentally, it is easier to use a phase-preserving amplifier instead of the assumed phase-sensitive amplifier with time-varying amplified quadrature. All our results remain the same for a phase-preserving amplifier, except the singular contribution to $K_{ff^*}$ becomes twice as large (in a real experiment the singular contribution broadens because of the finite bandwidth of the amplifier).

{\it Conclusions.}  We have developed the theory for analyzing the squeezing of a propagating microwave field in the transient regime. The most natural way to characterize squeezing in this case is via the two-time correlators $K_{\varphi_1\varphi_2}(t_1,t_2)$ of the detector output with different quadrature angles $\varphi_1$ and $\varphi_2$, since in experiments these correlators are directly related to the fluctuations of the integrated signal. In our theory the correlators $K_{\varphi_1\varphi_2}$ are expressed via the field fluctuation correlators $K_{ff}$ and $K_{ff^*}$, for which the differential equations have been derived using the semiclassical model. Our theory is equally applicable to squeezing in optics, though it is more challenging to realize transients of optical squeezing experimentally.

We thank Aashish Clerk, Justin Dressel, Farid Khalili, and Mark Dykman for useful discussions. The work was supported by ARO Grant No. W911NF-15-1-0496.

%\bibliography{References-squeezing}

\newpage

% Merge with supplemental materials
\clearpage %clear page
\onecolumngrid %one column
\vspace{\columnsep}
\begin{center}
\textbf{\large Supplemental Material for} \\
\vspace{0.2cm}
 \textbf{\large ``Two-time correlators for propagating squeezed microwave in transients''}
\end{center}
\vspace{\columnsep}
\twocolumngrid

%set counters to zero
\setcounter{equation}{0}
\setcounter{figure}{0}
\setcounter{table}{0}
\setcounter{page}{1}

%prepend S to eq. numbers
\renewcommand{\theequation}{S\arabic{equation}}
\renewcommand{\thefigure}{S\arabic{figure}}
\renewcommand{\bibnumfmt}[1]{[S#1]}
\renewcommand{\citenumfont}[1]{S#1}

%trick hyperref to avoid confusion
\renewcommand{\theHtable}{Supplement.\thetable}
\renewcommand{\theHfigure}{Supplement.\thefigure}
\renewcommand{\theHequation}{Supplement.\theequation}

\begin{center}
{\bf A. Correlators for propagating squeezed field:
 semiclassical model}
\end{center}

In this section we discuss a semiclassical description of the quantum fluctuations of the propagating field leaking out of the cavity. The main advantage of this approach is that it enables us to calculate the temporal correlators of the measured quadrature signals using classical stochastic equations. The semiclassical model is derived from the conventional input-output formalism~\cite{S-GardinerBook,S-Clerk2010}. We show that for a linear resonator, the correlators calculated within the semiclassical model are {\it exactly} equal to the correlators calculated in the standard quantum way.

\subsubsection{Correlators in the input-output formalism}

In the standard quantum approach, the correlator $K_{\varphi_1\varphi_2}(t_1, t_2)$ for the measured normalized quadrature signal $f_{\varphi}(t)$ with a time-varying quadrature phase $\varphi (t)$ is given by the symmetrized combination
    \begin{align}
K_{\varphi_1\varphi_2}^{\rm q}(t_1,t_2) =\frac{ \langle F_{\varphi_1}(t_1)\, F_{\varphi_2}(t_2)\rangle + \langle F_{\varphi_2}(t_2)\, F_{\varphi_1}(t_1) \rangle}{2} ,
    \label{eq:Suppl-corr-q}\end{align}
where $F_\varphi (t)$ is the quadrature operator for the propagating field.  It is related to the field operator $F(t)$ as
\begin{align}
\label{eq:Fphi-def-Suppl}
F_\varphi(t) = \frac{1}{2} \left[ e^{-i\varphi} F(t)+ e^{i\varphi} F^\dagger(t) \right] ,
\end{align}
and the Heisenberg picture is used for all operators.

In the input-output theory \cite{S-GardinerBook}, the field leaked from the cavity is written as
    \begin{align}
\label{eq:F-def-Suppl}
F(t)=-V(t) + \sqrt{\kappa}\, a(t) ,
    \end{align}
where $a(t)$ is the annihilation operator for the intracavity mode, for simplicity we assume $\kappa = \kappa_{\rm out}$, and the operator $V(t)$ of the incoming vacuum noise satisfies the commutation relations
    \be
    [V(t), V^\dagger (t')] =\delta (t-t') , \,\,\,  [V(t), V (t')] =0,
    \label{eq:V-comm-Suppl}\ee
while the average values of the products are
    \be
\langle V^\dagger(t)\, V (t')\rangle = \bar{n}_{\rm b}\, \delta (t-t') , \,\,\,  \langle V(t)\, V (t')\rangle =0,
    \label{eq:V-aver-Suppl}\ee
where $\bar{n}_{\rm b}=[\exp(\omega_{\rm r}/T)-1]^{-1}$ depends on the bath temperature $T$. In this section, we will assume $T=0$ ($\bar n_{\rm b}=0$), but generalization to a non-zero temperature is rather straightforward. The evolution of the operator $a(t)$ is~\cite{S-GardinerBook,S-Clerk2010}
    \be
    \dot{a}(t) =-\frac{\kappa}{2}\, a(t) +i[H(t), a(t)] +\sqrt{\kappa}\, V(t).
    \label{Suppl-a-dot}\ee

It is possible to show \cite{S-CarmichaelBook} that the propagating field $F(t)$ satisfies the same commutation relations as $V(t)$,
\begin{align}
\label{eq:corr-F-prop-Suppl}
[F(t),F^\dagger(t')]=\delta(t-t'), \,\,\, [F(t),F(t')]=0,
\end{align}
and, therefore, the correlator (\ref{eq:Suppl-corr-q}) for quadratures can be written without symmetrization,
\begin{align}\label{eq:corr-def-Suppl}
K_{\varphi_1\varphi_2}^{\rm q}(t_1,t_2) = \langle F_{\varphi_1}(t_1)\, F_{\varphi_2}(t_2)\rangle. %- \langle F_{\varphi_1}(t_1)\rangle\, \langle F_{\varphi_2}(t_2)\rangle.
\end{align}

Using Eq.~\eqref{eq:Fphi-def-Suppl}, we can write the correlator as
\begin{align}
\label{eq:Fphi1-Fphi2-corr-Suppl}
&\langle F_{\varphi_1}(t_1)\, F_{\varphi_2}(t_2)\rangle =  \frac{1}{4}\Big[ \langle F(t_1)\, F(t_2)\rangle \, e^{-i(\varphi_1+\varphi_2)}
\nonumber \\
& \hspace{0.5cm} + \langle F(t_1)\, F^\dagger(t_2)\rangle \, e^{i(\varphi_2 - \varphi_1)} +
\langle F^\dagger(t_1)\, F(t_2)\rangle \, e^{i(\varphi_1 - \varphi_2)}
    \nonumber \\
&\hspace{0.5cm}  + \langle F(t_2)\, F(t_1)\rangle^* e^{i(\varphi_1 + \varphi_2)}\Big],
\end{align}
where the two-time averages are  \cite{S-GardinerBook}
\begin{subequations}\label{eq:FFcorr--system-corr-Suppl}
\begin{align}
& \langle F(t_1)\, F(t_2)\rangle = \kappa \, \langle \mathcal{T}[a(t_2)\,a(t_1)]\rangle,
    \label{eq:FFcorr--system-corr-Suppl-1} \\
& \langle F(t_1)\, F^\dagger(t_2)\rangle = \delta(t_1-t_2) + \kappa \, \langle a^\dagger(t_2)\, a(t_1)\rangle,
    \label{eq:FFcorr--system-corr-Suppl-2} \\
& \langle F^\dagger(t_2)\, F(t_1)\rangle = \kappa \, \langle a^\dagger(t_2)\, a(t_1)\rangle.
\label{eq:FFcorr--system-corr-Suppl-3}\end{align}
\end{subequations}
The relations \eqref{eq:FFcorr--system-corr-Suppl-1}--\eqref{eq:FFcorr--system-corr-Suppl-3} are the standard results of the input-output theory; they are valid for arbitrary $t_1$ and $t_2$. In Eq.~\eqref{eq:FFcorr--system-corr-Suppl-1}, the time-ordering operator $\mathcal{T}$ is defined in the usual way: $\mathcal{T}[A(t_1)B(t_2)] = B(t_2)A(t_1)$ if $t_1<t_2$ and  $\mathcal{T}[A(t_1)B(t_2)] =A(t_1)B(t_2)$  if $t_1>t_2$.

Without loss of generality we assume $t_1<t_2$. Then, using Eqs.~\eqref{eq:corr-def-Suppl}--\eqref{eq:FFcorr--system-corr-Suppl}, the  correlator  (\ref{eq:Suppl-corr-q}) for the measured quadrature signal can be written as
\begin{align}
\label{eq:Kphi1phi2-simplified-Suppl}
&K^{\rm q}_{\varphi_1\varphi_2}(t_1,t_2) = \frac{\kappa}{2} \,
{\rm Re} \Big[\langle a(t_2)\, a(t_1)\rangle e^{-i(\varphi_1+\varphi_2)}
 \nonumber \\
&\hspace{3.2cm}
 + \langle a^\dagger(t_2)\, a(t_1)\rangle e^{-i(\varphi_1- \varphi_2)} \Big].
\end{align}

\subsubsection{Intracavity correlators via Wigner representation}

The two-time averages $\langle a(t_2)\, a(t_1)\rangle$ and $\langle a^\dagger(t_2)\, a(t_1)\rangle$ in Eq.\ (\ref{eq:Kphi1phi2-simplified-Suppl}) can be calculated using the standard  quantum-regression formulas~\cite{S-GardinerBook}
\begin{subequations}\label{eq:QR-formulas-Suppl}
\begin{align}
\label{eq:QR-formulas-Suppl-1}
& \langle a(t_2)\, a(t_1)\rangle = {\rm Tr}[a\, \tilde \rho(t_2)],
    \\ \label{eq:QR-formulas-Suppl-2}
& \langle a^\dagger(t_2)\,a(t_1)\rangle = {\rm Tr}[a^\dagger \tilde \rho(t_2)],
\end{align}
\end{subequations}
where $\tilde \rho(t)$ is an unphysical (in particular, non-Hermitian) density matrix, which has the initial condition
    \be
\tilde \rho(t_1) = a\, \rho(t_1),
    \label{eq:tilde-rho-in}\ee
relating it to the physical density matrix $\rho(t)$ of the resonator at the moment $t_1$, while in between $t_1$ and $t_2$ it evolves
in the same way as $\rho$  [cf.\   Eq.~\eqref{Suppl-a-dot}],
\begin{align}
\label{eq:SE-rho-Suppl}
\dot{\rho} = -i[H, \rho] + \kappa \, (a \rho a^\dagger - a^\dagger a \rho/2 - \rho a^\dagger a/2),
\end{align}
so that the equation for $\dot{\tilde\rho}$ is Eq.\ (\ref{eq:SE-rho-Suppl}) with $\rho$ replaced by $\tilde\rho$.
Note that in Eqs.\ (\ref{eq:QR-formulas-Suppl}) the left-hand sides assume  the Heisenberg picture, while the right-hand sides use the Schr\"odinger picture.

The justification of our semiclassical model for the output field can be based on the Wigner representation of the resonator density matrix $\rho(t)$. Instead of the standard Wigner function, depending on $x$ and $p$ [corresponding to  $(a+a^\dagger)/2$ and $(a-a^\dagger)/2i$],  we will use a slight modification, as in Refs.\ \cite{S-GardinerBook, S-CarmichaelBook}, in which the Wigner function depends on $\alpha =x+ip$ and $\alpha^*=x-ip$. So we will use the Wigner transformation $\mathcal{W}$ defined as
\begin{align}
\label{eq:Wigner-def}
&  \mathcal{W}[\rho (t)]= W(\alpha,\alpha^*,t)
    \nonumber \\
& \hspace{0.8cm} = \int  {\rm Tr}\big[\rho(t)\exp (z a^\dagger - z^*a)\big] \, e^{z^*\alpha - z \alpha^*} \, \frac{d^2 z}{\pi^2},
\end{align}
where $d^2 z \equiv d({\rm Re}\,z)\, d({\rm Im}\,z)$ corresponds to the integration over the complex phase space. Note that $W$ is real if $\rho$ is Hermitian (then complex conjugation of $W$ reduces to the transformation $z\to -z$). However, $W$ is complex if $\rho$ is non-Hermitian. The normalization following from Eq.\ \eqref{eq:Wigner-def} is $\int W(\alpha, \alpha^*, t)\, d^2\alpha = {\rm Tr}[\rho(t)]$. Note that the definition \eqref{eq:Wigner-def} allows us to think of $\alpha$ and $\alpha^*$ as independent variables, which are not necessarily conjugate to each other (even though the final expressions are evaluated for conjugate values). For example, the partial derivative $\partial_\alpha W\equiv \partial W/\partial \alpha$ is given by Eq.\ \eqref{eq:Wigner-def} with extra factor $z^*$ inside the integral, while $\partial_{\alpha^*} W$ produces the factor $-z$ inside the integral.

To use the Wigner representation in Eq.\ (\ref{eq:QR-formulas-Suppl}), we need to apply the Wigner transformation~\eqref{eq:Wigner-def} to the non-physical density matrix $\tilde\rho$; for that we will need the relations \cite{S-GardinerBook} (which are straightforward to derive)
\begin{align}
\label{eq:Wigner-identities-Suppl}
& \mathcal{W}[a\rho] = \Big( \alpha + \frac{1}{2}\partial_{\alpha^*}\Big) \, W, \,\,\,\, \mathcal{W}[\rho a^\dagger] = \Big( \alpha^* + \frac{1}{2}\partial_{\alpha}\Big) \, W,
    \nonumber \\
& \mathcal{W}[a^\dagger\rho] = \Big( \alpha^* - \frac{1}{2}\partial_{\alpha}\Big) \, W, \,\,\,\, \mathcal{W}[\rho a] = \Big( \alpha - \frac{1}{2}\partial_{\alpha^*}\Big) \, W.
\end{align}
Then using Eqs.~\eqref{eq:QR-formulas-Suppl},  \eqref{eq:tilde-rho-in}, and \eqref{eq:Wigner-identities-Suppl}, we can express the two-time averages in Eq.~\eqref{eq:Kphi1phi2-simplified-Suppl} as~\cite{S-GardinerBook}
\begin{subequations}\label{eq:aa-corr-Suppl}
\begin{align}
\label{eq:aa-corr-Suppl-1}
& \hspace{-0.1cm} \langle a(t_2) \, a(t_1)\rangle =
 \int  d^2\alpha_1 d^2\alpha_2
  \nonumber \\
& \hspace{1.7
cm}  \times \Big( \alpha_2 + \frac{1}{2}\partial_{\alpha_2^*} \Big) \, W(2|1) \Big( \alpha_1 + \frac{1}{2}\partial_{\alpha_1^*}\Big) \, W(1)
     \nonumber \\
&\hspace{0.3cm} =\int  \alpha_2\,W(2|1) \Big( \alpha_1 + \frac{1}{2}\partial_{\alpha_1^*}\Big)\, W(1) \, d^2\alpha_{1}d^2\alpha_2  ,
    \\
& \hspace{-0.2cm} \langle a^\dagger(t_2)\,a(t_1)\rangle  =
 \! \int \! \alpha_2^* W(2|1) \Big( \alpha_1 + \frac{1}{2}\partial_{\alpha_1^*}\Big) W(1) \, d^2\alpha_{1}d^2\alpha_2  , \label{eq:aa-corr-Suppl-2}
\end{align}
\end{subequations}
where $W(1)\equiv W(\alpha_1,\alpha_1^*,t_1)$ is the Wigner function at time $t_1$ and $W(2|1)\equiv W(\alpha_2,\alpha_2^*,t_2|\alpha_1,\alpha_1^*,t_1)$ is the propagator for the Wigner function from time $t_1$ to time $t_2$, which can be obtained from Eq.\ (\ref{eq:SE-rho-Suppl}). Note that the first and second forms of Eq.\ \eqref{eq:aa-corr-Suppl-1} differ only by the term $(1/2)\partial_{\alpha_2^*}$, which gives zero contribution after integration by parts over the whole space of $\alpha_2$; the same cancellation is used in Eq.\ \eqref{eq:aa-corr-Suppl-2}.

To find the propagator $W(2|1)$, we need to convert the evolution equation (\ref{eq:SE-rho-Suppl}) into the Wigner representation. The conversion is relatively simple for a linear resonator. Let us consider the rotating-frame Hamiltonian
    \begin{align}
H = \Omega \, a^\dagger a + \frac{i}{4}\big[{\varepsilon^* {a}^2 - \varepsilon a^\dagger}^2 \big] + \varepsilon_{\rm c}^* a  +\varepsilon_{\rm c} a^\dagger ,
    \label{eq:Hamiltonian-2}\end{align}
which is more general than the Hamiltonian (1) in the main text due to addition of a coherent drive with amplitude $\varepsilon_{\rm c}$, while the parametric drive still has the amplitude $\varepsilon$. For example, $\varepsilon_{\rm c}$ can represent the input signal in a parametric amplifier. Note that in the case $\varepsilon_{\rm c}\neq 0$, the intracavity and outgoing fields include the signal component, in contrast to the main text, where we considered only noise.
Also note that all parameters in the Hamiltonian ($\Omega$, $\varepsilon$, $\varepsilon_{\rm c}$) as well as the decay rate $\kappa$ can depend on time (this dependence should be slow in comparison with the resonator frequency $\omega_{\rm r}$, but can be arbitrarily fast compared with evolution in the rotating frame).

Applying the Wigner transformation \eqref{eq:Wigner-def} to Eq.\ (\ref{eq:SE-rho-Suppl}) and using Eqs.\ (\ref{eq:Wigner-identities-Suppl}) and (\ref{eq:Hamiltonian-2}), we obtain the following evolution equation for the Wigner function \cite{S-GardinerBook,S-CarmichaelBook}
\begin{align}
\label{eq:Wigner-evol}
& \partial_{t}W = -\partial_{\alpha} (\Lambda W)  -\partial_{\alpha^*} (\Lambda^* W) +\frac{\kappa}{2} \, \partial^2_{\alpha\alpha^*}W ,
    \\
& \Lambda(\alpha,\alpha^*,t)=-  \left( \frac{\kappa}{2}+i\Omega \right)\alpha - \frac{\varepsilon}{2}\, \alpha^* -i \varepsilon_{\rm c} .
\label{eq:Lambda-Suppl}\end{align}
Note that in this derivation, the relations \eqref{eq:Wigner-identities-Suppl} should be applied several times, e.g.,
\begin{align}
& \mathcal{W}[a^2\rho] = \Big( \alpha + \frac{1}{2}\partial_{\alpha^*}\Big)^ 2  W,
\,\,\,
\mathcal{W}[\rho a^2] = \Big( \alpha - \frac{1}{2}\partial_{\alpha^*}\Big)^2 W,
    \nonumber \\
& \mathcal{W}[a^\dagger a\rho] = \Big( \alpha^* - \frac{1}{2}\partial_{\alpha}\Big) \Big( \alpha + \frac{1}{2}\partial_{\alpha^*}\Big) \, W.
\end{align}

Most importantly, Eq.\ \eqref{eq:Wigner-evol} for the Wigner function evolution has the same form as the Fokker-Planck equation \cite{S-RiskenBook,S-GardinerBook} for the evolution of a probability distribution, in which
$ \Lambda(\alpha,\alpha^*,t)$ has the physical meaning of a drift velocity. Moreover, Eq.\ \eqref{eq:Lambda-Suppl} for $\Lambda$ has the same form as for the evolution of a classical field in the cavity.
This similarity between the Wigner function and the classical probability distribution will be the basis of the proof that the correlator \eqref{eq:Suppl-corr-q} can be calculated within the semiclassical model.

Note that for the averages \eqref{eq:QR-formulas-Suppl} we need to consider the evolution of an unphysical (non-Hermitian) density matrix $\tilde\rho$ and therefore unphysical (complex) Wigner function $W$. Nevertheless,  for the propagator $W(2|1)$ in Eqs.\ \eqref{eq:aa-corr-Suppl}, it is sufficient to consider physical (real) $W$ in Eq.\ (\ref{eq:Wigner-evol}). This is because of the linearity of Eq.\ \eqref{eq:SE-rho-Suppl} (linearity of quantum mechanics), so that $W(2|1)$ is just the Green's function of Eq.\  \eqref{eq:Wigner-evol} with initial condition $W(\alpha_2,\alpha_2^*,t_1|\alpha_1,\alpha_1^*,t_1)=\delta^2 (\alpha_1-\alpha_2)\equiv \delta\big({\rm Re} (\alpha_2-\alpha_1)\big)\, \delta\big({\rm Im}(\alpha_2-\alpha_1)\big)$.

Moreover, in the calculation of the propagator $W(2|1)$ using Eq.\ (\ref{eq:Wigner-evol}), the Wigner function $W$ remains positive, since the initial condition $\delta^2 (\alpha_1-\alpha_2)$ is positive and Eq.~\eqref{eq:Wigner-evol} is a second-order partial differential equation. (This fact follows from Pawula's theorem \cite{S-RiskenBook}; normalization for the Wigner function is preserved automatically.) Therefore, in this case the Wigner function $W$ can be interpreted as a classical probability distribution in phase space, which evolves due to the drift $\Lambda$ and diffusion (see below). Similarly, $W(1)$ in Eq.\ \eqref{eq:aa-corr-Suppl} is also positive (and therefore can be interpreted as a classical probability distribution) if the Wigner function at some earlier time $t_0<t_1$ is positive. This is what we assume below; for example, assuming that in a distant past the evolution started from vacuum (which has positive Wigner function).

\subsubsection{Semiclassical model}

As discussed in the main text, in the semiclassical model we consider a stochastic evolution of the classical field $\alpha (t)$ in the cavity, which is caused by the Hamiltonian, dissipation, and classical complex-valued noise $v(t)$, which imitates the vacuum noise $V(t)$ incident on the cavity from the transmission line. This noise has correlators
\begin{align}
\label{eq:noise-corr-Suppl}
\hspace{-0.0cm} \langle v (t)\, v ^*(t')\rangle = (\bar{n}_{\rm b}+1/2)\,\delta (t-t'), \,\,\, \langle v (t)\, v (t')\rangle =0,
\end{align}
which are classical counterparts of the quantum relations \eqref{eq:V-comm-Suppl} and \eqref{eq:V-aver-Suppl}. For simplicity we assume zero temperature, so that $\bar{n}_{\rm b}=0$, though generalization to a non-zero temperature is simple.

The evolution of the intracavity field [counterpart of Eq.\ \eqref{Suppl-a-dot}] is
    \begin{align}
    \dot\alpha = -\frac{\kappa}{2}\, \alpha -i \partial_{\alpha^*} h(\alpha,\alpha^*) +\sqrt{\kappa} \, v(t) ,
    \label{eq:alpha-dot-Suppl-1}\end{align}
where $h(\alpha, \alpha^* )$ is the classical Hamiltonian, corresponding to the quantum Hamiltonian $H$. For the Hamiltonian  (\ref{eq:Hamiltonian-2}) of a driven {\it linear} resonator, the classical Hamiltonian $h$ can be obtained from $H$ by simply replacing $a$ with $\alpha$ and $a^\dagger$ with $\alpha^*$, so that
    \be
    h(\alpha, \alpha^*) = \Omega |\alpha|^ 2 + \frac{i}{4} \left[ \varepsilon^* \alpha^2 -\varepsilon (\alpha^*)^2 \right] + \varepsilon_{\rm c}^* \alpha +\varepsilon_{\rm c} \alpha^*,
    \ee
and therefore the field evolution is
    \begin{align}
    \dot\alpha = -\left(\frac{\kappa}{2} +i\Omega\right) \alpha -\frac{\varepsilon}{2}\, \alpha^*- i\varepsilon_{\rm c} +\sqrt{\kappa} \, v(t) .
    \label{eq:alpha-dot-Suppl-2}\end{align}
In general, the classical Hamiltonian $h$ should be chosen so that $\dot{\alpha}$ correctly describes the evolution of the field $\alpha$ in the classical case.
An initial condition for $\alpha(t)$ is usually not needed, because if the evolution starts at $t=-\infty$, then the initial condition does not matter. However, if we want to start evolution from $t=t_0$, then $\alpha(t_0)$ in the semiclassical model should be treated as a random complex number, with two-dimensional probability distribution equal to the Wigner function $W(\alpha, \alpha^*,t_0)$, which is positive (and normalized) by the above-discussed assumption.

The outgoing field [counterpart of Eq.\ \eqref{eq:F-def-Suppl}] is
    \be
\label{eq:f-def-Suppl}
    f(t) = -v(t) +\sqrt{\kappa} \, \alpha(t),
    \ee
and the measured quadrature signal $f_\varphi (t)$ for the quadrature phase $\varphi$ is
    \be
\label{eq:fphi}
    f_\varphi(t) =\frac{1}{2}\left[ e^{-i\varphi} f(t) +e^{i\varphi} f^*(t) \right] .
    \ee

\vspace{0.2cm}

Our goal is to prove that the two-time correlator $K_{\varphi_1\varphi_2}^{\rm q}(t_1, t_2)$ for the signal $f_\varphi (t)$ calculated in the quantum way (\ref{eq:corr-def-Suppl}) is {\it exactly} equal to the correlator
\begin{align}
\label{eq:corr-class-Suppl}
&K_{\varphi_1\varphi_2}(t_1,t_2) = \langle f_{\varphi_1}(t_1)\, f_{\varphi_2}(t_2) \rangle,
%- \langle f_{\varphi_1}(t_1)\rangle \, \langle f_{\varphi_2}(t_2) \rangle,
\end{align}
calculated in the semiclassical model. Note that here $\alpha (t)$ and $f_\varphi(t)$ contain contributions due to ``signal'' $\alpha_{\rm c} (t)$, in contrast to the main text, where we considered only noise (in the main text $\alpha_{\rm c}=0$).

Let us first check the equivalence for the singular contribution to the correlator at $t_1=t_2$ (then $\varphi_1=\varphi_2$ as well). From Eqs.\ \eqref{eq:Fphi-def-Suppl}--\eqref{eq:V-aver-Suppl} we see that at zero temperature the singular part of the quantum correlator is $(1/4)\, \delta (t_1-t_2)$, and from Eqs.\ \eqref{eq:noise-corr-Suppl} and \eqref{eq:f-def-Suppl}--\eqref{eq:corr-class-Suppl} we obtain the same result for the semiclassical correlator. Next, for the equivalence in the case $t_1\neq t_2$, it is sufficient to consider $t_1<t_2$ (because of the symmetry). In this case the quantum correlator is given by Eq.\ (\ref{eq:Kphi1phi2-simplified-Suppl}), while the semiclassical correlator is %
\begin{align}
\label{eq:corr-class-2-Suppl}
&K_{\varphi_1\varphi_2}(t_1,t_2) = \frac{1}{2} \,
{\rm Re}\big[\langle f(t_2) f(t_1)\rangle\,e^{-i(\varphi_1 + \varphi_2)}
 \nonumber \\
 & \hspace{2.9cm} + \langle f^*(t_2)f(t_1)\rangle\, e^{-i(\varphi_1 - \varphi_2)} \big].
\end{align}
Therefore, we only need to prove two relations for $t_1<t_2$:
\begin{align}
%\label{eq:proof-1-Suppl}
%\langle f(t)\rangle =&\, \sqrt{\kappa}\,\langle a(t)\rangle, \\
\label{eq:proof-2-Suppl}
& \langle  f(t_2) \, f(t_1) \rangle = \kappa \, \langle a(t_2)\, a(t_1)\rangle, \\
& \langle f^*(t_2)\, f(t_1) \rangle = \kappa \, \langle a^\dagger(t_2)\, a(t_1)\rangle .  \label{eq:proof-3-Suppl}
\end{align}

Let us prove Eq.~\eqref{eq:proof-2-Suppl} first [the proof of Eq.~\eqref{eq:proof-3-Suppl} is similar].
Using Eq.\ \eqref{eq:f-def-Suppl} for $f(t)$, we can write the left-hand side of Eq.~\eqref{eq:proof-2-Suppl} as
\begin{align}
\langle f(t_2) f(t_1)\rangle = \kappa\, \langle \alpha(t_2) \, \alpha(t_1)\rangle - \sqrt{\kappa}\, \langle \alpha(t_2)\, v(t_1) \rangle ,
\end{align}
since $\langle v(t_2)\, \alpha(t_1) \rangle=0$ for $t_1<t_2$. Comparing this equation with Eq.~\eqref{eq:aa-corr-Suppl-1}, we see that we can prove Eq.~\eqref{eq:proof-2-Suppl} by proving the following two relations:
\begin{align}
\label{eq:alpha-alpha-proof}
& \langle \alpha(t_2) \, \alpha(t_1)\rangle = \int \alpha_2 \, W(2|1) \, \alpha_1 \, W(1) \, d^2\alpha_{1}d^2\alpha_2,
    \\
\label{eq:alpha-v-proof}
&  \frac{-2}{\sqrt{\kappa}}\, \langle\alpha(t_2) \, v(t_1)\rangle =\int \alpha_2 \, W(2|1) \, \partial_{\alpha_1^*}W(1) \, d^2\alpha_{1}d^2\alpha_2.
\end{align}
Note that in these relations, $\alpha(t)$ in the left-hand side is the semiclassical random process, while $\alpha_1$ and $\alpha_2$ in the right-hand side are the integration variables.

To prove Eq.\ \eqref{eq:alpha-alpha-proof}, let us show that the Wigner function $W(\alpha,\alpha^*,t)$ is equal to the probability distribution of $\alpha (t)$ in the semiclassical model. Introducing the probability distribution $P(x,p,t)$ on the two-dimensional plane with real coordinates $x={\rm Re}(\alpha )$ and $p={\rm Im}(\alpha )$, from the Langevin equation  \eqref{eq:alpha-dot-Suppl-2} with noise given by Eq.\ \eqref{eq:noise-corr-Suppl}, we can write the standard Fokker-Planck equation
    \begin{align}
& \partial_t P = -\partial_x \{ [-(\frac{\kappa}{2} +{\rm Re} \frac{\varepsilon}{2})x +(\Omega -{\rm Im} \frac{\varepsilon}{2}) p +{\rm Im}\, \varepsilon_{c}] P  \}
    \nonumber \\
& \hspace{0.9cm} -\partial_p \{ [-(\frac{\kappa}{2} -{\rm Re} \frac{\varepsilon}{2})p -(\Omega +{\rm Im} \frac{\varepsilon}{2}) x -{\rm Re}\, \varepsilon_{c}] P \}
    \nonumber \\
& \hspace{0.9cm} + \frac{\kappa}{8} (\partial_x^2 +\partial_p^2) P.
    \label{eq:FP-1-Suppl}\end{align}
It is easy to check that if we formally introduce the same probability distribution as a function of $\alpha$  and $\alpha^*$, i.e., $P(x,p,t)= \tilde{P} (\alpha,\alpha^*, t)$, then Eq.\ \eqref{eq:FP-1-Suppl} can be rewritten as
    \begin{align}
& \partial_t \tilde{P} = -\partial_\alpha \{ [-(\kappa /2 + i\Omega) \alpha - (\varepsilon /2) \alpha^* -i \varepsilon_{c}] \tilde{P}  \}
    \nonumber \\
& \hspace{0.9cm} -\partial_{\alpha^*} \{ [-(\kappa /2 - i\Omega) \alpha - (\varepsilon^* /2) \alpha +i \varepsilon_{c}^*] \tilde{P}  \}
    \nonumber \\
& \hspace{0.9cm} + (\kappa /2) \, \partial_{\alpha\alpha^*}^2 \tilde{P}.
    \label{eq:FP-2-Suppl}\end{align}
This is exactly the same equation as Eq.\ \eqref{eq:Wigner-evol} for the Wigner function. Therefore, if $W(\alpha, \alpha^*, t_0)= P(x,p,t_0)$ at some initial time $t_0$ (as we assumed above), then the Wigner function will be equal to the probability distribution of $\alpha(t)$ in the semiclassical model at any later time, $W(\alpha, \alpha^*, t)= P(x,p,t)$.

Thus, we have shown that $W(1)$ in Eq.\ \eqref{eq:alpha-alpha-proof} is equal to the probability distribution of the semiclassical intracavity field $\alpha$ at time $t_1$. Similarly, the propagator $W(2|1)$ in Eq.\ \eqref{eq:alpha-alpha-proof} is equal to the probability distribution of the field $\alpha(t_2)$ at time $t_2$ in the semiclassical model if at time $t_1<t_2$ the field is $\alpha_1$. Therefore,  Eq.~\eqref{eq:alpha-alpha-proof} is obviously valid.

It is a little more difficult to prove Eq.\ \eqref{eq:alpha-v-proof}. Let us introduce discrete time with very small time steps $\Delta t\to 0$. Then the average $\langle \alpha (t_2)\, v(t_1)\rangle $ in the left-hand side of Eq.\ \eqref{eq:alpha-v-proof} is replaced with $\langle \alpha (t_2)\, \tilde{v}(t_1)\rangle$, where $\tilde v(t_1)=(1/\Delta t)\int_{t_1}^{t_1+\Delta t} v(t)\, dt$. Now $\tilde{v}(t_1)$ is a (large) complex number, which is Gaussian-distributed with $\langle |\tilde{v}|^2\rangle =1/(2\Delta t)$ and $\langle \tilde{v}\rangle =0$. Because of the linearity of the Fokker-Planck equation,
    \begin{align}
\langle \alpha (t_2)\, \tilde{v}(t_1)\rangle = \int \alpha_2 P(2|1) \, \langle \delta P(1) \, \tilde{v}(t_1)\rangle \, d^2\alpha_1 d^2\alpha_2,
    \label{eq:alpha-v-2}\end{align}
where $P(2|1)$ is the propagator for probabilities [the same as $W(2|1)$, we used a different notation only to emphasize that we consider the semiclassical model] and $\delta P(1)\equiv P(x_1+\delta x_1, p_1+\delta p_1, t_1+\Delta t)-P(x_1, p_1, t)$ is the change of the probability distribution between time moments $t_1$ and $t_1+\Delta t$ due to the kick to $\alpha(t)$ produced by $\tilde v(t_1)$. Note that the averaging in the right-hand side of Eq.\ (\ref{eq:alpha-v-2}) is only over $\tilde v(t_1)$, while in the left-hand side it also includes averaging over random trajectories.  As seen from Eq.\ \eqref{eq:alpha-dot-Suppl-2}, the noise $\tilde v(t_1)$ shifts $\alpha(t_1)$ by $\delta \alpha(t_1)=\sqrt{\kappa}\, \tilde v(t_1)\,\Delta t$, and therefore the leading-order change of the probability distribution is $\delta P(1) = -\sqrt{\kappa}\, [\partial_x P \, {\rm Re} \, \tilde v(t_1)+ \partial_p P \, {\rm Im} \, \tilde v(t_1)]\, \Delta t$. Now using averages $\langle \tilde v (t_1) \, {\rm Re}\, \tilde v (t_1) \rangle \, \Delta t = 1/4$ and $\langle \tilde v (t_1) \, {\rm Im}\, \tilde v (t_1) \rangle \, \Delta t = i/4$, we obtain
    \begin{align}
& \langle \alpha (t_2)\, \tilde{v}(t_1)\rangle = -\frac{\sqrt{\kappa}}{4} \int \alpha_2 P(2|1) \, [\partial_{x_1} P(1) + i \partial_{p_1}P(1)] \,
    \nonumber \\
&\hspace{3.5cm} \times d^2\alpha_1 d^2\alpha_2.
    \end{align}
This equation is the same as Eq.\ \eqref{eq:alpha-v-proof} since $W(2|1)=P(2|1)$ and $\partial_{\alpha_1^*} W(1)= (1/2)\, \partial_{x_1} P(1) +(i/2)\, \partial_{p_1} P(1)$ [as follows from the change of variables: $x=(\alpha+\alpha^*)/2$ and $p=(\alpha-\alpha^*)/2i$]. Thus,  we have proved Eq.\ \eqref{eq:alpha-v-proof}.

By proving Eqs.\ \eqref{eq:alpha-alpha-proof} and \eqref{eq:alpha-v-proof}, we have proved Eq.\ \eqref{eq:proof-2-Suppl}. The proof of Eq.\  \eqref{eq:proof-3-Suppl} is very similar. Instead of Eqs.\  \eqref{eq:alpha-alpha-proof} and \eqref{eq:alpha-v-proof}, we need to prove relations
 \begin{align}
\label{eq:alpha*-alpha-proof}
& \langle \alpha^*(t_2) \, \alpha(t_1)\rangle = \int \alpha_2^* \, W(2|1) \, \alpha_1 \, W(1) \, d^2\alpha_{1}d^2\alpha_2,
    \\
\label{eq:alpha*-v-proof}
&  \frac{-2}{\sqrt{\kappa}}\, \langle\alpha^*(t_2) \, v(t_1)\rangle =\int \alpha_2^* \, W(2|1) \, \partial_{\alpha_1^*}W(1) \, d^2\alpha_{1}d^2\alpha_2,
\end{align}
which can be done in the same way as above.

Thus, we have shown that for a system with the Hamiltonian (\ref{eq:Hamiltonian-2}), the quantum correlators \eqref{eq:Suppl-corr-q} are {\it exactly} equal to the correlators \eqref{eq:corr-class-Suppl} calculated within the semiclassical model.

\vspace{0.5cm}
\begin{center}
{\bf B. Transient squeezing in a weakly nonlinear resonator}
\end{center}

In the main text we considered the case of a linear resonator, with squeezing produced by a parametric drive. However, the main initial motivation for this work was to develop a theory capable of calculating fluctuations of the integrated signal in circuit QED measurement of a
superconducting qubit (these fluctuations are directly related to the probability of error in qubit measurement). In the process of qubit measurement, the squeezing is self-generated due to resonator nonlinearity induced by interaction with the qubit \cite{S-Khezri2016,S-Khezri2017}. Therefore, it is important to consider the nonlinear case as well.

As shown in Ref.\ \cite{S-Khezri2017}, in the case of a weakly nonlinear resonator (typical in qubit measurement), the intracavity state remains approximately Gaussian during the transient squeezing caused by abruptly applied coherent drive. This is why the theory developed in the main text is directly applicable to this case as well. However, while for a linear resonator our semiclassical theory is exact, for a weakly nonlinear resonator it is only approximate. (We believe it is a very good approximation for typical parameters of a qubit measurement; however, the accuracy still has to be analyzed.)

In a weakly nonlinear case, the Hamiltonian can be written as
\begin{align}
H = \sum\nolimits_n E(n)|n\rangle \langle n| + \frac{i}{4}\big[{\varepsilon^* {a}^2 - \varepsilon a^\dagger}^2 \big] + \varepsilon_{\rm c}^* a  +\varepsilon_{\rm c} a^\dagger ,
    \label{eq:Hamiltonian-nonlin}\end{align}
where $|n\rangle$ is $n$th eigenstate of the resonator and the rotating-frame energy $E(n)$ is related to the weakly changing (lab-frame) resonator frequency $\omega_{\rm r}(n)$ as
    \be
    E(n)= \sum_{k=0}^{n-1} [\omega_{\rm r}(n)-\omega_{\rm rf}].
    \ee
Here $\omega_{\rm rf}$ is an (arbitrary) rotating frame frequency, for which it is most natural to choose the frequency of the coherent drive. A small difference between the frequencies of the rotating frame, coherent drive, and (halved) parametric drive can be taken into account by slowly changing  complex amplitudes $\varepsilon_{\rm c}$ and $\varepsilon$. Note that for the qubit measurement we usually do not need parametric drive, $\varepsilon=0$, but we keep this term for generality.

The main idea of using the semiclassical model in this case is to separate evolution of the average field $\alpha_{\rm c}$ (corresponding to the maximum of the Wigner function, which is assumed to be Gaussian) and fluctuations $\delta\alpha$,
    \begin{align}
\label{eq:alpha-decomp}
\alpha(t) = \alpha_{\rm c}(t) + \delta \alpha(t),
    \end{align}
so that the ``center'' evolves as
    \be
    \dot{\alpha}_{\rm c}=
      -i \left[\omega_{\rm r} (|\alpha_{\rm c}|^2) -\omega_{\rm rf}\right] \alpha_{\rm c}  -\frac{\kappa}{2}\, \alpha_{\rm c} -\frac{\varepsilon}{2}\, \alpha^*_{\rm c}- i\varepsilon_{\rm c}  ,
    \label{eq:alpha-dot-center} \ee
while for fluctuations we use linearization \cite{S-Ludwig1975, S-CarmichaelBook} near the center,
   \begin{align}
& \frac{d}{dt}\,\delta\alpha = - i \left[ \omega_{\rm r}(|\alpha_{\rm c}|^2)-\omega_{\rm rf}+ \frac{d\omega_{\rm r}}{dn}\,  |\alpha_{\rm c}|^2\right] \delta\alpha - \frac{\kappa}{2}\, \delta\alpha
    \nonumber \\
&\hspace{1.2cm}  -\left( \frac{\varepsilon}{2}+i \frac{d\omega_{\rm r}}{dn} \, \alpha_{\rm c}^2  \right)\, (\delta\alpha)^*   +\sqrt{\kappa} \, v(t) ,
    \label{eq:alpha-dot-fluct}\end{align}
where $d\omega_{\rm r}(n)/dn$ is evaluated at $n=|\alpha_{\rm c}|^2$. Note that the terms with $d\omega_{\rm r}/dn$ come from the contribution $-i\alpha_{\rm c} (d\omega_{\rm r}/dn) \, 2{\rm Re} (\alpha_{\rm c}^* \, \delta\alpha)$ describing the resonator frequency change due to fluctuations.

Equation (\ref{eq:alpha-dot-fluct}) has the same form as Eq.\ (6) in the main text, with  $\alpha$ replaced by $\delta\alpha$ (in the main text we considered only fluctuations) and also $\Omega$ and $\varepsilon$ replaced by the corresponding terms in Eq.\ (\ref{eq:alpha-dot-fluct}). Therefore, squeezing of fluctuations in the case of a weakly nonlinear resonator is still described by Eqs.\ (16)--(18) of the main text ($\alpha \to\delta\alpha$, $f \to \delta f\equiv f-\sqrt{\kappa}\, \alpha_{\rm c}$), with the following replacements in Eq.\ (14) of the main text:
  \begin{align}
& \Omega \, \to \,  \omega_{\rm r} (|\alpha_{\rm c}|^2) -\omega_{\rm rf} + \left. \frac{d\omega_{\rm r}(n)}{dn} \right|_{n=|\alpha_{\rm c}|^2} |\alpha_{\rm c}|^2 ,
    \\
&    \varepsilon \,\, \to \, \varepsilon + 2 i \left. \frac{d\omega_{\rm r}(n)}{dn} \right|_{n=|\alpha_{\rm c}|^2} \alpha_{\rm c}^2 ,
    \end{align}
where $\alpha_{\rm c}$ depends on time via Eq.\ \eqref{eq:alpha-dot-center}.

\end{document}